\documentclass[aps,prl,twocolumn,superscriptaddress]{revtex4-2}
\usepackage[colorlinks=true,citecolor=blue,linkcolor=magenta]{hyperref}
\usepackage[utf8]{inputenc}
\usepackage[english]{babel}
\usepackage{amsmath}
\usepackage{graphicx}
\usepackage{xcolor}
\usepackage{amsfonts}
\usepackage{natbib}
\usepackage{subfigure}
\usepackage{caption}
\usepackage{ragged2e}
\usepackage{balance}
\newtheorem{theorem}{Theorem}[section]

\newtheorem{lemma}[theorem]{Lemma}

\begin{abstract}
The connection between the Maximum Entropy (MaxEnt) formalism and Restricted Boltzmann Machines (RBMs) is natural, as both give rise to a Boltzmann-like distribution with constraints enforced by Lagrange multipliers, which corresponds to RBM parameters. We integrate RBMs into quantum state tomography (QST) by using them as probabilistic models to approximate quantum states while satisfying MaxEnt constraints. Additionally, we employ polynomially efficient quantum sampling techniques to enhance RBM training, enabling scalable and high-fidelity quantum state reconstruction. This approach provides a computationally efficient framework for applying RBMs to MaxEnt-based quantum tomography. Furthermore, our method applies to the general and previously unaddressed case of reconstructing arbitrary mixed quantum states from incomplete and potentially non-commuting sets of expectations of observables while still ensuring maximal entropy.  
\end{abstract}

\begin{document}
\title{Maximal Entropy Formalism and the Restricted Boltzmann Machine}
\author{Vinit Singh}
\affiliation{Department of Electrical and Computer Engineering, North Carolina State University, Raleigh, NC 27606}
\author{Rishabh Gupta}
\affiliation{Department of Chemistry, Purdue University, West Lafayette, IN 47907, USA}
\author{Manas Sajjan}
\affiliation{Department of Electrical and Computer Engineering, North Carolina State University, Raleigh, NC 27606}

\author{Francoise Remacle}
\affiliation{Theoretical Physical Chemistry, UR MOLSYS, University of Liege, B4000 Liège, Belgium}
\author{Raphael D. Levine}
\affiliation{The Fritz Haber Center for Molecular Dynamics and Institute of Chemistry, The Hebrew University of Jerusalem, Jerusalem 91904}
\author{Sabre Kais}
\email{Corresponding  author: skais@ncsu.edu}
\affiliation{Department of Electrical and Computer Engineering, North Carolina State University, Raleigh, NC 27606}

\maketitle


\pagenumbering{gobble}

\section{Introduction}
Quantum State Tomography (QST) is a crucial procedure for reconstructing quantum states from measurement data, enabling the validation and characterization of quantum systems \cite{levine2009m,alhassid,komarova2020surprisal,komarova2021compacting,blavier2022entanglement}. Among the various approaches to QST, the Maximal Entropy (MaxEnt) formalism \cite{Jaynes,Jaynes2,raphy,raphy2} is particularly compelling due to its principled foundation: it provides the least biased estimation of a quantum state consistent with observed measurement statistics. This method ensures that no unwarranted assumptions are made about unmeasured aspects of the state, making it a natural choice for state reconstruction in scenarios where limited measurement data is available \cite{gupta2021,gupta2021convergence,gupta2022,makhija2024time}. 

In this work, we explore the use of Restricted Boltzmann Machines (RBMs) as an ansatz 
within the MaxEnt formalism to perform QST. A computationally efficient algorithm that retains the core principles of the MaxEnt method while overcoming any computational bottlenecks.  
A significant limitation of the standard MaxEnt formalism is the challenge of determining the Lagrange multipliers, 
For general cases, this involves the exponentiation of large matrices, which is exponentially costly in both computation and storage. 
This limitation makes MaxEnt-based QST intractable for large quantum systems. 
Algebraic methods \cite{hamilton2025evo} can alleviate some of this problem but face the challenge of non-commuting observables.
Our approach mitigates this issue by using an RBM as a variational ansatz for the quantum state. By leveraging the representational efficiency of RBMs, we ensure that our state reconstruction protocol remains computationally feasible, avoiding any exponential scaling in storage or computation.

Our framework provides a two-fold advantage: (i) the RBM ansatz enables efficient storage of quantum states using polynomial resources, and (ii) RBMs support efficient sampling-based optimization techniques, making parameter training computationally viable. Neural Quantum States (NQS), of which RBMs are a subclass, have been shown to effectively represent a broad class of quantum states, making them well-suited for our tomography protocol.\cite{sharir2022neural}


\begin{figure*}[ht!]
    \centering
    \includegraphics[width=1\linewidth]{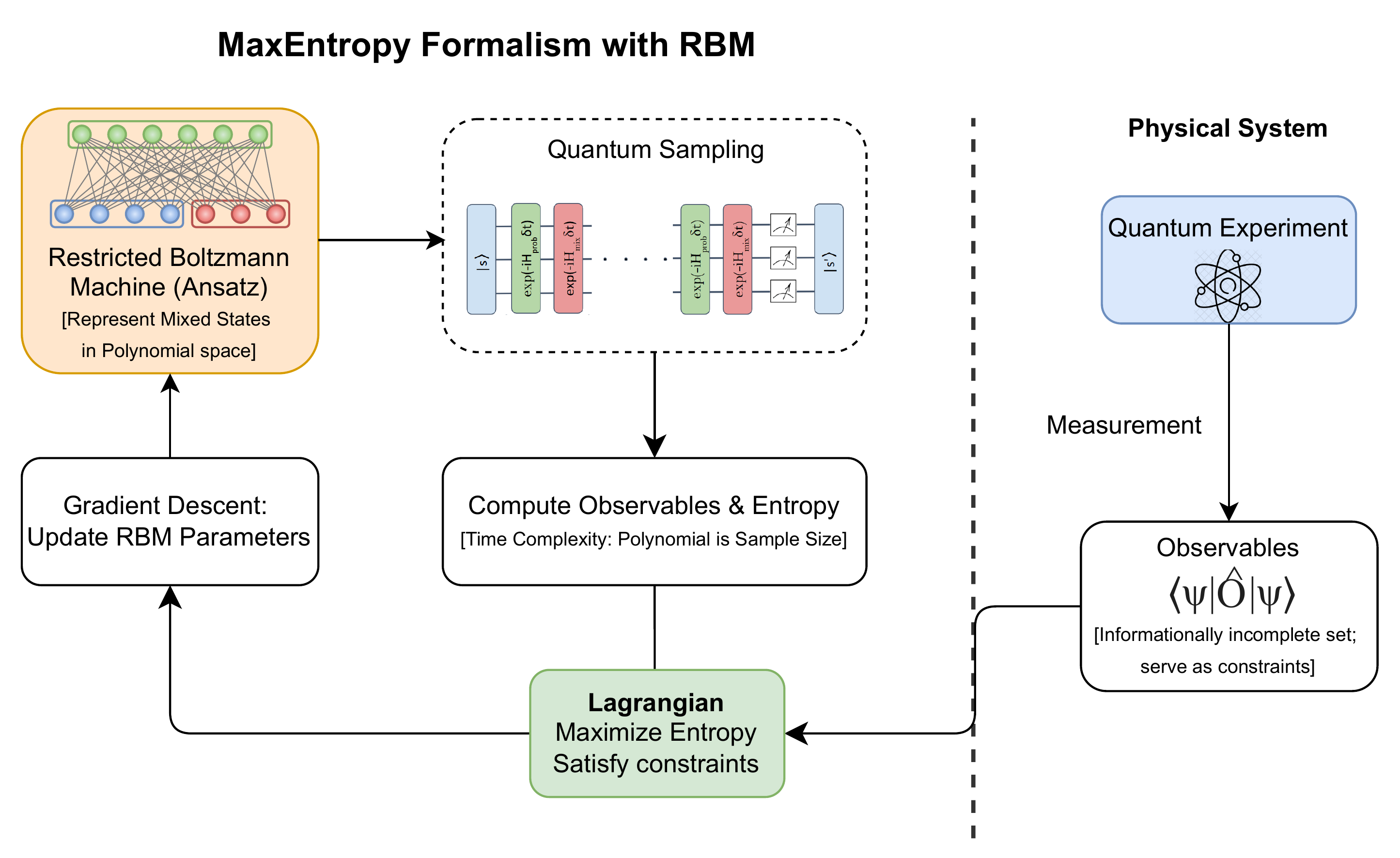}
    \caption{\justifying \textbf{A schematic illustrating the workflow for reconstructing quantum states using the Maximum Entropy (MaxEnt) formalism with a Restricted Boltzmann Machine (RBM).} The RBM serves as an ansatz for the state to be reconstructed; it requires only polynomial storage in the system size and supports an efficient, sampling-based optimization algorithm.
    The goal is to reconstruct the state of a physical quantum system from which measurements are obtained during a quantum experiment. These yield a set of observables, which may be informationally incomplete. These observables serve as constraints for the reconstruction process.
    Using the MaxEnt formalism, a Lagrangian is formulated to maximize the entropy of the quantum state, subject to the constraint that the expectation values of the observables match those derived from experimental data. The quantum state is represented using the RBM ansatz, which is capable of modeling mixed states with polynomial computational resources. The RBM parameters are optimized via gradient descent, iteratively refining the model to satisfy the MaxEnt constraints.
    The optimization cycle is depicted by the loop in the schematic: starting from the current RBM parameters, samples are generated using quantum hardware. These samples are then used to estimate observables and entropy (with a computational cost that scales polynomially with the number of samples), which are the key components in the MaxEnt Lagrangian. The Lagrangian is subsequently optimized using a gradient-based algorithm, thereby updating the RBM parameters.
    The final RBM-based state representation enables quantum sampling to simulate measurements or compute state properties. This hybrid approach—combining RBM with MaxEnt—offers an efficient and unbiased method for quantum state reconstruction, even with an informationally incomplete measurement set.}
    \label{fig:Schematic}
\end{figure*}

Our approach reconstructs a quantum state that satisfies the imposed measurement constraints, i.e. the reconstructed state reproduces the expected observables while maximizing the entropy of state that fulfills these conditions. Thus, it inherits the desirable properties of the MaxEnt formalism—yielding the least biased state possible—while achieving efficient storage and computation. Moreover, our method is completely general: it can reconstruct both pure and mixed quantum states from an arbitrary set of measurement observables, making it broadly applicable to various quantum systems.

Moreover, while classical sampling-based methods exist for training NQS in polynomial time and space, they are often slow. These computations can be further accelerated using quantum-assisted sampling techniques. Recent advancements in our group have led to the development of efficient quantum algorithms for evaluating and training RBMs on quantum hardware \cite{sajjan2024polynomially}, offering a new paradigm for scalable, high-fidelity quantum state reconstruction. By integrating these quantum-accelerated RBM techniques with the MaxEnt formalism, we propose a novel methodology for performing QST directly on a quantum computer. This approach not only enhances the computational feasibility of MaxEnt-based tomography but also provides a framework for quantum-native learning of quantum states.

The remainder of this paper is structured as follows: In the next section, we introduce Restricted Boltzmann Machines (RBM) and discuss how classical RBMs can be enhanced using quantum-assisted sampling techniques. We then provide a brief overview of the MaxEnt formalism and its relationship with RBMs. Then, we outline how RBMs can be integrated into the QST framework, ensuring a general and scalable protocol that remains efficient for arbitrary sets of measurement constraints. Finally, we present numerical simulations demonstrating the effectiveness of our method.

\newpage

\subsection{Restricted Boltzmann Machine} 

Restricted Boltzmann Machines (RBMs) are generative probabilistic models that represent complex distributions through a bipartite graphical structure consisting of visible and hidden units. As an ansatz for learning probability distributions, RBMs optimize a set of free parameters—weights and biases—to accurately model the underlying distribution. The network consists of two distinct layers: the visible layer, containing \( n \) neurons that encode the system state, and the hidden layer, consisting of \( m \) neurons that introduce additional degrees of freedom to enhance model expressivity.  

Mathematically, an RBM defines a joint probability distribution over visible and hidden binary random variables governed by an energy function analogous to that of a classical Ising model with partial connectivity. Given a system with \( n \) visible neurons \( \sigma_i \in \{1, -1\} \) and \( m \) hidden neurons \( h_j \in \{1, -1\} \), the energy function is given by  
\begin{equation}
    E(\vec{\sigma}, \vec{h}) = -\sum_i a_i \sigma_i - \sum_j b_j h_j - \sum_{ij} W_{ij} \sigma_i h_j,
\end{equation}

where \( a_i \) and \( b_j \) are the bias terms associated with the visible and hidden neurons, respectively, and \( W_{ij} \) represents the coupling strength between visible neuron \( i \) and hidden neuron \( j \). The probability distribution over the joint visible-hidden state follows the Gibbs distribution:  
\begin{equation}
    P(\vec{\sigma}, \vec{h}) = \frac{1}{Z} \exp(-E(\vec{\sigma}, \vec{h}))
    \label{rbm}
\end{equation}


where \( Z = \sum_{\vec{\sigma}, \vec{h}} \exp(-E(\vec{\sigma}, \vec{h})) \) is the partition function ensuring normalization. By marginalizing over the hidden variables, one obtains the effective probability distribution over the visible units, which serves as a powerful generative model for learning complex data distributions, including quantum states. The visible neurons encode the system state, while the hidden neurons introduce additional degrees of freedom that enhance the model’s expressivity, allowing it to capture higher-order correlations. Notably, the probability distribution obtained from the RBM ansatz is mathematically equivalent to the thermal (Gibbs) distribution of the Ising model at an effective temperature set by the model parameters. The statistical mechanics formulation of RBMs reveals a direct analogy with the Ising model: visible and hidden neurons correspond to spins, the weights \( W_{ij} \) act as interaction terms, and the biases \( a_i \) and \( b_j \) serve as external fields. This connection underscores the deep interplay between RBMs and statistical physics, making them particularly relevant for quantum state reconstruction.

\subsection{Quantum-Enabled Restricted Boltzmann Machines}

Restricted Boltzmann Machines (RBMs) have demonstrated remarkable expressivity in approximating complex quantum states. However, their classical training relies on Markov Chain Monte Carlo (MCMC) sampling, which suffers from slow convergence due to long autocorrelation times due to conventional transition proposals. Quantum-assisted sampling technique provides a pathway to accelerate this process and significantly enhance the efficiency of training neural-network quantum states.  
In our recent work, we developed a quantum-enabled variational Monte Carlo (Q-VMC) framework for efficiently training neural-network quantum states, particularly RBMs, using quantum-assisted sampling techniques.\cite{sajjan2024polynomially} The core idea is to construct a quantum circuit that generates samples from a variationally optimized surrogate probability distribution, which approximates the RBM’s target distribution. This is achieved by defining a surrogate Ising-like network with an equivalent Hamiltonian:  
\begin{equation}
    H_{\text{sur}} = \sum_i l_i \sigma_i^z + \sum_{i,j} J_{ij} \sigma_i^z \sigma_j^z.
\end{equation}

where the parameters \( l_i \) and \( J_{ij} \) are learned to best approximate the probability distribution of the RBM ansatz. The sampling process is then performed using a quantum circuit implementing a Trotterized evolution of the surrogate Hamiltonian,  
\begin{equation}
    U(\tau, \gamma) = e^{-i\gamma H_{\text{sur}} \tau} e^{-i(1-\gamma) H_x \tau}
\end{equation}

where \( H_x = \sum_i \sigma_i^x \) acts as a quantum mixing term. By tuning the evolution parameters \( \tau \) and \( \gamma \), this quantum sampling protocol efficiently explores the probability landscape, overcoming the limitations of classical MCMC approaches.  

The quantum-enabled Variational Monte Carlo (VMC) method for training NQS offers significant advantages over classical training methods:  

\begin{itemize}
    \item \textbf{Faster Convergence}: Quantum proposals (e.g., Trotterized circuits) exhibit a spectral gap $\delta$ that decays three times slower than classical MCMC proposals (local/uniform sampling), reducing mixing time and yielding less correlated samples. This accelerates convergence to the target distribution.
    \item \textbf{Reduced Variance}: The quantum-assisted sampler achieves five times lower \(l_2\)-norm error in distribution approximation compared to classical methods, enhancing sample quality. Energy variance is low and is further reduced by zero-variance extrapolation (ZVE) to achieve $<0.5\%$ relative energy error. 
    \item \textbf{Resource Efficiency}: The quantum circuit scales linearly in number of qubits (\(O(n)\)) and depth of quantum circuit (\(O(\tau n)\)), with \(O(N_s)\) circuit executions independent of system size. Trotterized circuits use \(O(n^2)\) two-qubit gates per layer, parallelized to \(O(n)\) depth, making them feasible for near-term quantum hardware.
\end{itemize}

By integrating this quantum-assisted sampling into the training, we achieve a robust and scalable algorithm for learning quantum states using RBMs.

\subsection{Maximum Entropy Formalism}

The Maximum Entropy (MaxEnt) formalism \cite{katz1967principles, levine1985statistical} reconstructs a quantum state \(\hat{\rho}_{ME}\) that both satisfies constraints imposed by measured observables and maximizes the Rényi entropy. The Lagrangian governing the optimization is formulated in terms of the density matrix \(\hat{\rho}\) and a set of observables \(\hat{O}_i\):

\begin{equation}
L = S_{\alpha}(\hat{\rho}) + \lambda_0 \left[\text{Tr}(\hat{\rho}) -1\right] + \sum_k \lambda_k \left[\text{Tr}(\hat{\rho} \hat{O}_k) - \langle \hat{O}_k \rangle \right].
\end{equation}

Here, the constraints ensure that the reconstructed state reproduces the measurement statistics of the target state. In particular, the constraint \(\hat{O}_0 = I\) enforces the normalization condition \(\text{Tr}(\hat{\rho}) = 1\). 

In the standard literature, the von Neumann entropy is most commonly used. Maximizing the Lagrangian analytically with respect to \(\hat{\rho}\) leads to the well-known thermal (Gibbs) state:
\[
\hat{\rho}_{ME} = Z^{-1} \exp\left(-\sum_{k=1} \lambda_k \hat{O}_k\right),
\]
where the partition function \(Z = \exp(\lambda_0)\) ensures normalization, and the \(\lambda_k\) are Lagrange multipliers enforcing the constraints, i.e., \( \text{Tr}(\hat{\rho}_{ME} \hat{O}_k) = \langle\hat{O}_k\rangle \).

The advantage of standard maximal entropy has been amply demonstrated with special reference to a few particle systems far away from equilibrium \cite{levine2009m, levine1985statistical, levine1988chemical}. Determining a quantum mechanical density matrix of maximal entropy brings these advantages to the age of quantum technologies \cite{gupta2021, gupta2022, hamilton2024constructing} and time-evolving systems \cite{makhija2024time, tishby1984time, Hamilton}.

Despite its success, the standard MaxEnt formalism has significant limitations when applied to general quantum systems. One major limitation is that it requires the observables \(\hat{O}_k\) to commute. This restriction arises because the density matrix must be Hermitian and positive semi-definite, which in turn requires that it commutes with the quantum surprisal,  
\[
\hat{I} = -\sum_{k=0} \lambda_k \hat{O}_k.
\]  
This condition holds only when all observables \(\hat{O}_k\) mutually commute, which is often not the case in real quantum experiments. As a result, the standard MaxEnt approach is only applicable to a restrictive subset of observables, limiting its practicality for general quantum state reconstruction.  
Another significant challenge in the MaxEnt formalism is determining the Lagrange multipliers \(\lambda_k\), which, for most general cases, requires exponential storage and computation. 

Our approach addresses these issues by introducing an RBM-based ansatz for the mixed-density matrix. Unlike the state for standard MaxEnt $\hat{\rho}_{ME}$, our RBM-based ansatz is inherently Hermitian and positive semi-definite by construction. This is achieved through a purification technique, where a wavefunction is defined over an extended Hilbert space, and the mixed state is obtained by tracing out auxiliary (environment) qubits. Since the ansatz guarantees the necessary properties of the density matrix, we are not restricted to using only commuting observables. This makes our approach significantly more general and applicable to arbitrary quantum systems.  

Instead of explicitly solving for the Lagrange multipliers, in our approach, we directly optimize the Lagrangian numerically using an \textbf{augmented Lagrangian technique}. This involves defining a modified cost function such that its minimization leads to the maximization of entropy and enforcing the constraints through a penalty term. 
\begin{equation}
C = - S(\rho) + \sum_i \xi_i \Big[\langle \hat{O}_i \rangle_\rho - O_i^{target} \Big]^2
\label{lagrange_opt}
\end{equation}
The constraints are controlled by hyperparameters \(\xi_k\), which adjust the strictness of the constraint enforcement. Unlike Lagrange multipliers, these hyperparameters do not need to be precisely determined for the optimization to be effective, greatly simplifying the process. In practice, The $\xi$'s are kept small initially to ensure that the cost function doesn't explode as the estimated values of operators could be far from the target; as the training progresses, the values of $\xi$'s increased gradually to ensure that the constraints are tight and the entropy is maximized accordingly.

Another key difference in our approach is the use of $\alpha$-R{\'e}nyi entropy instead of von Neumann entropy (VNE). There are two primary reasons for this choice: (a) R{\'e}nyi entropy can be efficiently computed using sampling techniques, compared to VNE. \cite{wang2020calculating} (b) Higher-order R{\'e}nyi entropies (\(S_\alpha\) for \(\alpha \geq 2\)) provide a lower bound on VNE. This means that maximizing \(S_2\) indirectly maximizes the VNE. Therefore, even when working with VNE, most of the optimization can be performed using \(S_2\), with only fine-tuning using VNE estimates towards the end of training is sufficient. 

By addressing these fundamental limitations, our method provides a flexible and computationally efficient alternative to the standard MaxEnt formalism. It allows for the use of non-commuting observables and eliminates the need to solve for Lagrange multipliers.



\subsection{Relationship between RBM and MaxEntropy}

One can show the relationship between Restricted Boltzmann Machines (RBMs) and the ansatz used in MaxEntropy formalism. Both ansatzes share the same fundamental origin as they describe thermal states defined over specific energy functions. In the case of RBMs, the energy function is given by a parameterized interaction between visible and hidden units. In the MaxEnt framework, the energy function is effectively determined by the constraints imposed in the Lagrangian. 

The equivalence emerges by choosing \( O_k \) as Pauli-\(Z\) operators corresponding to the RBM’s visible and hidden units. Specifically, the RBM density operator satisfies the constraints:  
\begin{align}
    \text{Tr}(\hat{\rho}_{RBM} &\hat{\sigma}_{zi}^v) = \langle \hat{\sigma}_{zi}^v \rangle, \nonumber \quad
    \text{Tr}(\hat{\rho}_{RBM} \hat{\sigma}_{zj}^h) = \langle \hat{\sigma}_{zj}^h \rangle, \nonumber \quad \\
    &\text{Tr}(\hat{\rho}_{RBM} \hat{\sigma}_{zi}^v \hat{\sigma}_{zj}^h) = \langle \hat{\sigma}_{zi}^v \hat{\sigma}_{zj}^h \rangle.
\end{align}

Comparing with the MaxEnt formulation, the Lagrange multipliers \( \{ \lambda_k \} \) map directly onto the RBM parameters:  
\[
\lambda_i \leftrightarrow a_i, \quad \lambda_j \leftrightarrow b_j, \quad \lambda_{ij} \leftrightarrow W_{ij}.
\]  

This mapping establishes RBMs as an implicit realization of the MaxEnt principle, where the RBM parameters act as the driving forces that encode statistical constraints. 


\section{Integrating RBM into the QST Framework}
Building upon the equivalence between RBM and MaxEnt formalism, we perform quantum state tomography (QST) by using RBM as a probabilistic model for quantum states. Tracing out the hidden units of RBM yields a probability distribution over the visible neurons, providing an ansatz that approximates the quantum state while satisfying measurement constraints. Since a quantum state can have complex coefficients, in general, the parameters of RBM ($\theta = \{ a, b, W\}$) are kept complex, and the distribution over the visible neurons is derived by marginalizing the hidden neurons from the joint RBM distribution. Equation (\ref{rbm}) defines the full RBM state, while summing over the hidden units leads to the description of the quantum state $|\psi\rangle$:  
\begin{eqnarray}
    |\psi_\theta (\vec{\sigma})\rangle &=&  \frac{1}{Z}{\sum_{\vec{h}} \exp\left[{\sum_i a_i \sigma_i + \sum_j b_j h_j + \sum_{ij} W_{ij} \sigma_i h_j}\right]}  \nonumber \\
    &=&  \frac{1}{Z}{\exp\left({\sum_i a_i \sigma_i}\right) \prod_{j=1}^m 2\cosh{(b_j + \sum_i W_{ij} \sigma_i)} }.  \label{rbm_red}
\end{eqnarray}  

This formulation shows that the effective probability distribution over the visible units retains an RBM structure but without explicit dependence on hidden variables. By utilizing this ansatz, we construct a quantum state that adheres to the constraints imposed by MaxEnt while benefiting from the RBM’s representational power. This approach allows us to bypass the strict structural limitations of the original RBM–MaxEnt equivalence, making it a more flexible and scalable alternative for quantum state reconstruction.



\begin{figure}[h!]
    \centering
    \includegraphics[width=\linewidth]{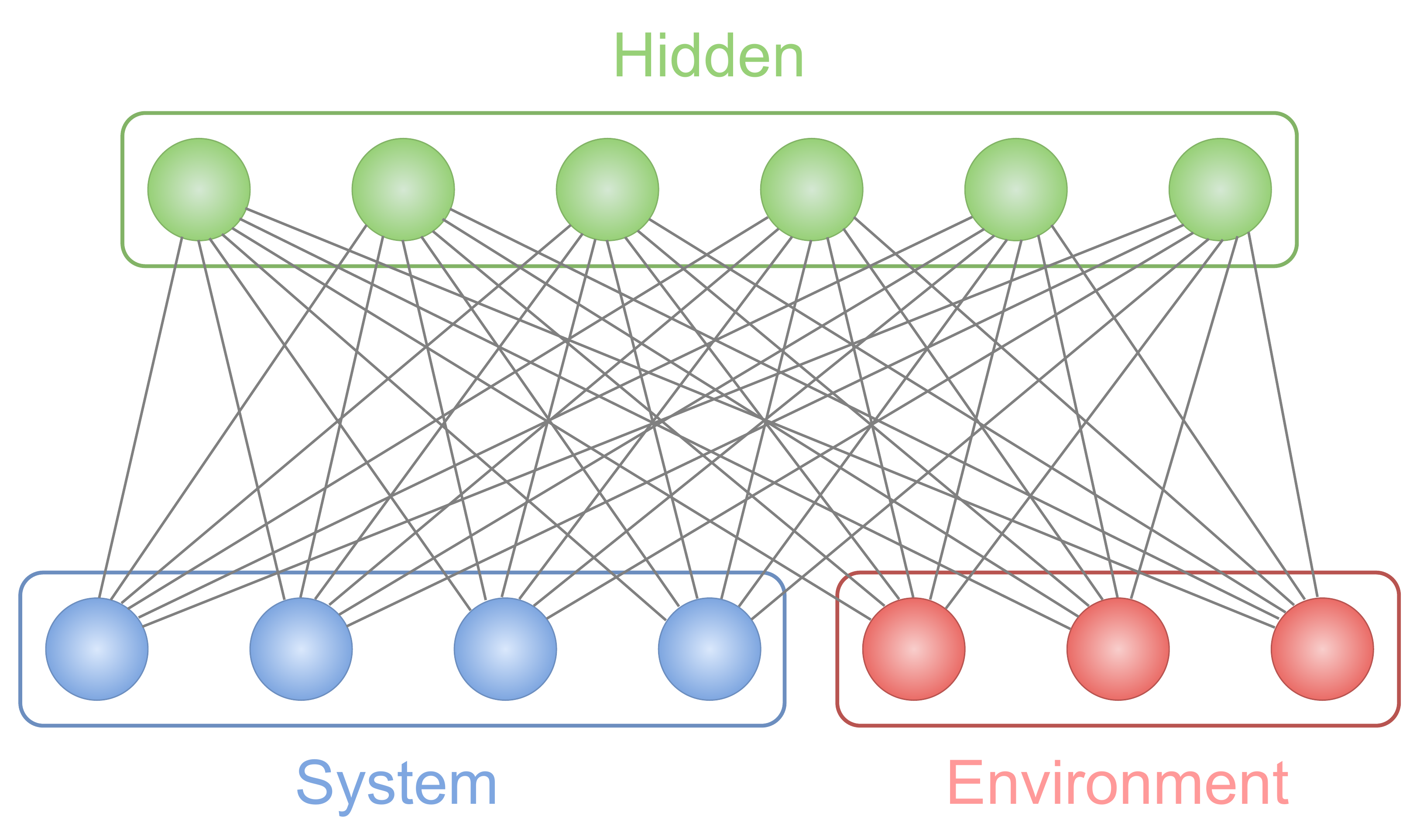}
    \caption{\justifying Graphical representation of the Restricted Boltzmann Machine (RBM) used to represent a mixed density matrix. The visible layer (blue) encodes the state of the physical system, while the other two layers are used to describe the mixing due to the environment (red) and to capture the hidden correlations between the physical degrees of freedom (green).}
    \label{fig:Mixed_RBM}
\end{figure}

This formalism can be generalized to a density matrix, and our quantum-enabled algorithm can efficiently handle this case. To obtain a mixed-density matrix of the system, we would need to extend the size of the state by appending some additional degrees of freedom. We would have a larger pure state, which, when ancillas are traced out, leaves behind a mixed state for the system. In figure [\ref{fig:Mixed_RBM}] the composite state $\rho_\theta^{\sigma \oplus a}$ is pure, and therefore $\rho_\theta^{\sigma \oplus a} = |\psi_\theta\rangle  \langle\psi_\theta|$, with a neural network wave function $|\psi_\theta\rangle = \sum_{\sigma a} \psi_\theta (\sigma,a) |\sigma\rangle \otimes |a\rangle$. The density matrix is simply obtained by tracing out the auxiliary system $\rho_\theta  = Tr_a \{|\psi\rangle \langle\psi_\theta| \}$, 
\[\rho_\theta(\sigma, \sigma') = \sum_{a} \psi_\theta (\sigma,a) \psi^*_\theta (\sigma',a)  \]



\paragraph{\textbf{Observable estimation}:}
To estimate the expectation values of the observables \( \hat{O} \), we sample from the probability distribution \(\rho^v_v\) using Markov Chain Monte Carlo (MCMC) methods. Our quantum-enabled algorithm accelerates this process by reducing the number of samples required to evaluate the expectation values. The expectation value of an observable \(\hat{O}\) is computed as:

\begin{align}
    \langle \hat{O} \rangle &= Tr(\rho \hat{O}) \nonumber = \sum_v \rho_v^v \Big( \frac{\sum_{v'} \rho_{v'}^v O_v^{v'}}{\rho_v^v} \Big) \nonumber\\
    &= \sum_v \rho_v^v O_{loc}(v) = \langle O_{loc}(v) \rangle_{\rho_v^v} \label{Eq. S1}
\end{align}
    
where \(O_{\mathrm{loc}}(v) = \frac{\sum_{v'} \rho_{v'}^v O_v^{v'}}{\rho^v_v}\) is the local observable at state v. The estimation of $O_{loc}(v)$ is conventionally done explicitly under the assumption that the operator $O$ is sparse and for a given configuration $v$, the connected configurations $v'$ scale only polynomially with system size.

\paragraph{\textbf{Entropy estimation}} The most challenging part in the optimization of the Lagrangian in the MaxEnt formalism is entropy estimation. Though in the standard MaxEnt formalism, Von Neumann entropy (VNE) is maximized, computing VNE is relatively difficult compared to other R{\'e}nyi entropies. \cite{wang2020calculating}
Since we just need to maximize the entropy of the system, and we know that the second-order R{\'e}nyi entropy ($S_2(\rho)$) lower bounds VNE, maximizing $S_2(\rho)$, in turn, maximizes $S_1(\rho)$. So, in our work, we maximize the $S_2(\rho)$ as it's relatively cheaper to compute to optimize the Lagrangian.

Here, we explore some of the sampling-based techniques to estimate various R{\'e}nyi entropies for NQS building on the works by \cite{wang2020calculating}. Estimating second order R{\'e}nyi entropy is relatively straightforward, as it can be obtained by estimating the SWAP operator \cite{hastings2010measuring} 

\[ S_2(\rho_A) =  -\log [Tr(\rho_A^2)] = -\log (\langle \psi \otimes \psi |  SWAP_A | \psi \otimes \psi \rangle)\]

Now since we are working with a mixed density matrix which is derived from a pure state in the extended Hilbert space, estimating the R{\'e}nyi entropy at the partition between the system and environment qubits will give us the desired entropy of the mixed state for the system. In the above equation, the SWAP operator acts as shown in the figure \ref{fig:SWAP}:

\begin{figure}
    \centering
    \includegraphics[width=1\linewidth]{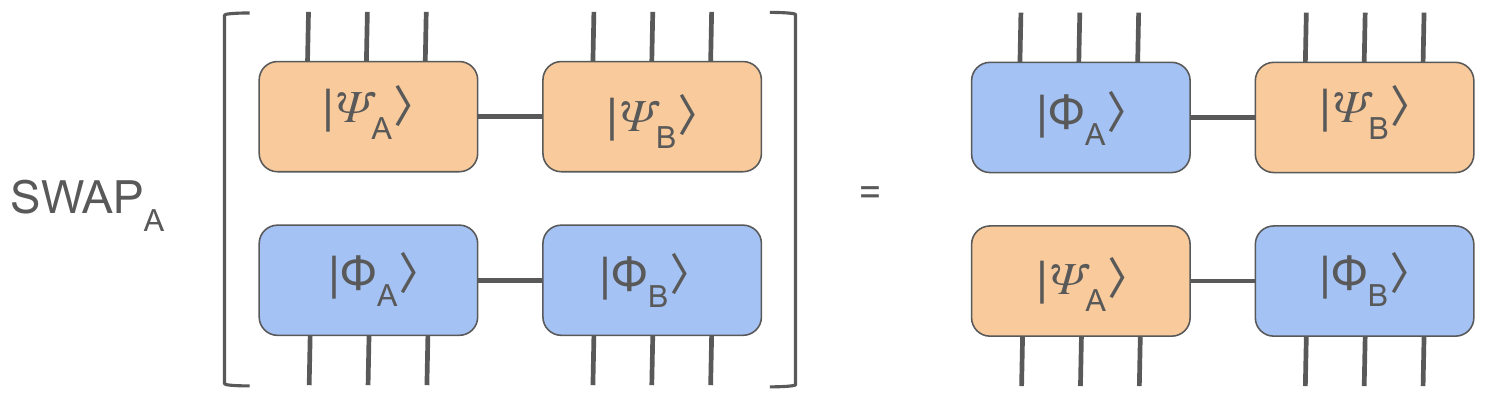}
    \caption{\justifying A tensor network representation of the action of SWAP operator over a product of two states ($|\psi\rangle \otimes |\phi\rangle$) across the region A.}
    \label{fig:SWAP}
\end{figure}

Just like R{\'e}nyi entropy of second order, other higher orders can be estimated by using permutation operators as mentioned in \cite{wang2020calculating}. Estimation of Von Neumann entropy can be performed using the polynomial approximation :  

\begin{equation}
    S_1(\rho) = \text{Tr}(\rho \ln \rho) \approx \sum_{\alpha=1}^{n_c} \alpha_n Tr[\rho^n].
\end{equation}  

Where $n_c$ is the cutoff polynomial degree. The coefficients of the polynomial ($\alpha_n$) are described in the Appendix.  

Once entropy and observables are estimated, the Lagrangian could now be optimized to reconstruct the state following the MaxEntropy formalism. The optimization of entropy while following the constraints of the problem is done by the augmented Lagrangian cost function defined in Eq. [\ref{lagrange_opt}]


\subsection{Resource requirements}
Our approach makes use of both quantum and classical resources. First, samples are generated using quantum circuit executions, and then classical post-processing is used to estimate the desired expectation values and gradients. The number of samples, \( N_s \), required to estimate observables is determined by the desired precision and follows the relation \( N_s \approx O(\text{Var}(\hat{O})/\epsilon_{obs}^2) \). This highlights that the sampling complexity is directly influenced by the variance of the observable and the targeted accuracy.

The quantum circuit requirements for our model are primarily characterized by the qubit count, gate count, and circuit depth. The number of qubits required corresponds to the visible layer size (n) and is independent of the hidden layer size (m). The gate count and circuit depth primarily arise from Trotterization. The initial state preparation consists of \( O(n) \) single-qubit \( \sigma_x \) gates, which can be executed in parallel, leading to a depth of \( O(1) \). The Trotterization process introduces a more significant depth overhead, with each Trotter layer requiring \( O(n) \) single-qubit gates, \( O(n^2) \) two-qubit \( R_{zz} \) gates (equivalent to \( O(2n^2) \) CNOTs), and an overall depth of \( O(n) \) per layer. The total number of Trotter steps, \( N_{\text{trot}} \), determines the accuracy of the circuit. It is important to note that perfect accuracy is not necessary for generating samples, as these samples are subsequently filtered using the Metropolis-Hastings criterion. In practice, a reasonably noisy circuit with a finite number of Trotter steps is sufficient. For a more detailed analysis of the quantum resource requirements, see \cite{sajjan2024polynomially}.

Once the required samples are generated using quantum circuit executions for a given set of parameters pf RBM, estimating the expectation values of \( k \) observables requires \( O(k N_s) \) time. For entropy estimation, we generate samples from the state twice, incurring a time cost of \( O(2 N_s) \), and compute the \( SWAP_{\text{loc}} \) for each sample in \( O(1) \) time. Consequently, the second-order Rényi entropy computation scales as \( O(N_s) \), while higher-order \( \alpha \)-Rényi entropy calculations require \( O(\alpha N_s) \) time. The polynomial approximation of the von Neumann entropy (VNE) up to degree \(\alpha_{max}\) requires \( O(\alpha_{\max}^2 N_s) \) time. Here, the coefficients of the polynomials are precomputed and stored. The computation cost arises from the estimation of all the Renyi entropies up to order \(\alpha_{max}\) for VNE estimation. The max time required for cost function evaluation thus scales as \( O((\alpha_{\max}^2 + k) N_s) \). 

Gradient estimation benefits from the availability of analytical gradient expressions, allowing us to reuse the generated samples. Given that there are \( O(2(n+m+nm)) \) parameters in the RBM (accounting for real and complex components), the gradient computation scales as \( O(nm (\alpha_{\max}^2 + k) N_s) \). Therefore, overall, each training step incurs a total computational cost of 
\[\boxed{T_{epoch} =  O(nm (\alpha_{max}^2 + k) N_s T_{cl} + N_s T_q)} \]
where \( T_q \) is the time required for a single quantum circuit execution and $T_{cl}$ is the time required for a single unit of post-processing computation on classical hardware. The dominant factor in practice is the time required for quantum circuit execution, which is heavily dependent on the quality of the quantum device. All the classical postprocessing, i.e., computations of expectation values and gradients, can be heavily accelerated by parallelization, thereby making $T_{cl}$ small.

\begin{figure}[ht!]
  \centering 
\includegraphics[width=3.5in]{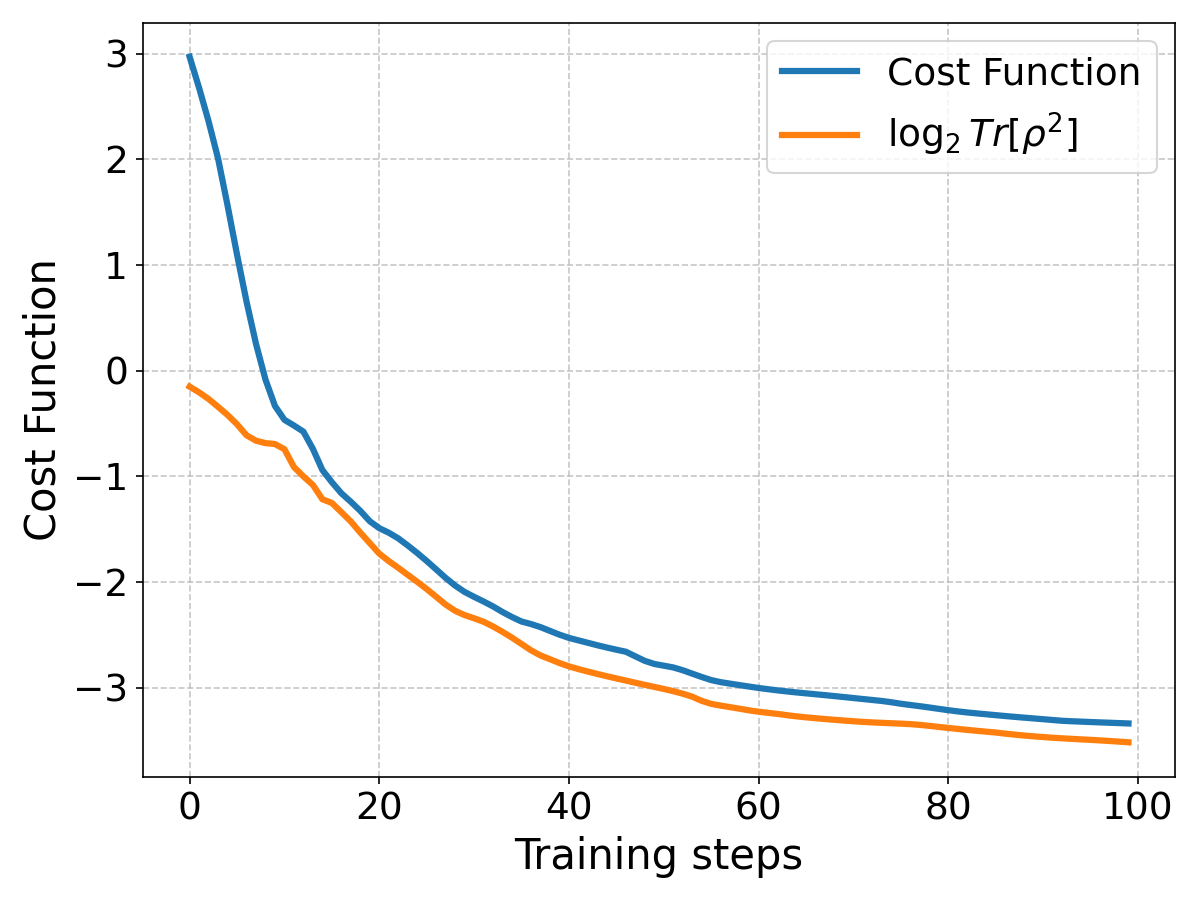} 
\caption{\justifying Progression of the median of cost function and negative of second order Renyi entropy [$\log_2 Tr[\rho^2]$] of the reconstructed quantum state during training. The median is computed over 10 experiments run for reconstructing 6 qubit mixed-states (system qubits = 6, environment qubits = 4, hidden units=4, number of observables = 6) }
\label{fig_training}
\end{figure} 

\begin{figure}[ht!]
\includegraphics[width=\linewidth]{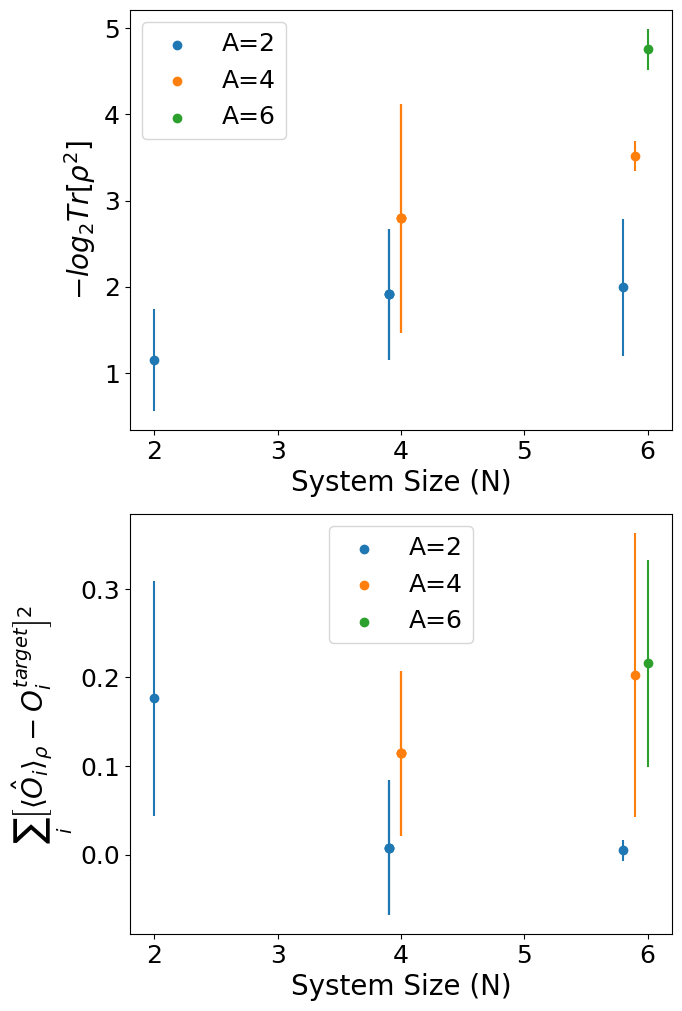} 
\caption{\justifying The figure depicts the final results obtained after training the RBM to reconstruct the given density matrix using Maximal Entropy Formalism. The figures shows (top) final entropy ($S_2(\rho) = -log_2 Tr[\rho^2]$) and (bottom) the observable constraint ($\langle \hat{O}_i \rangle_\rho - O_i^{target}$) obtained post-training 10 different state reconstruction experiments. Each state had a different set of target density matrices, known observables, and initial parameters for RBM. For every System Size ($N$), we also include the results for various Environment Sizes ($A \leq N$). The results from various environment sizes are shown by different colors clustered around the system size. The dashed black line in the entropy plot shows the upper bound of entropy for a given system size. Note: For experiments where the environment sizes ($A$) are less than the system size, the size of the environment qubit dictates the maximum amount of entanglement generated, i.e. $S_2 (\rho) \leq A$}
\label{fig_results}
\end{figure}

\begin{figure}[ht!]
  \centering 
\includegraphics[width=3in]{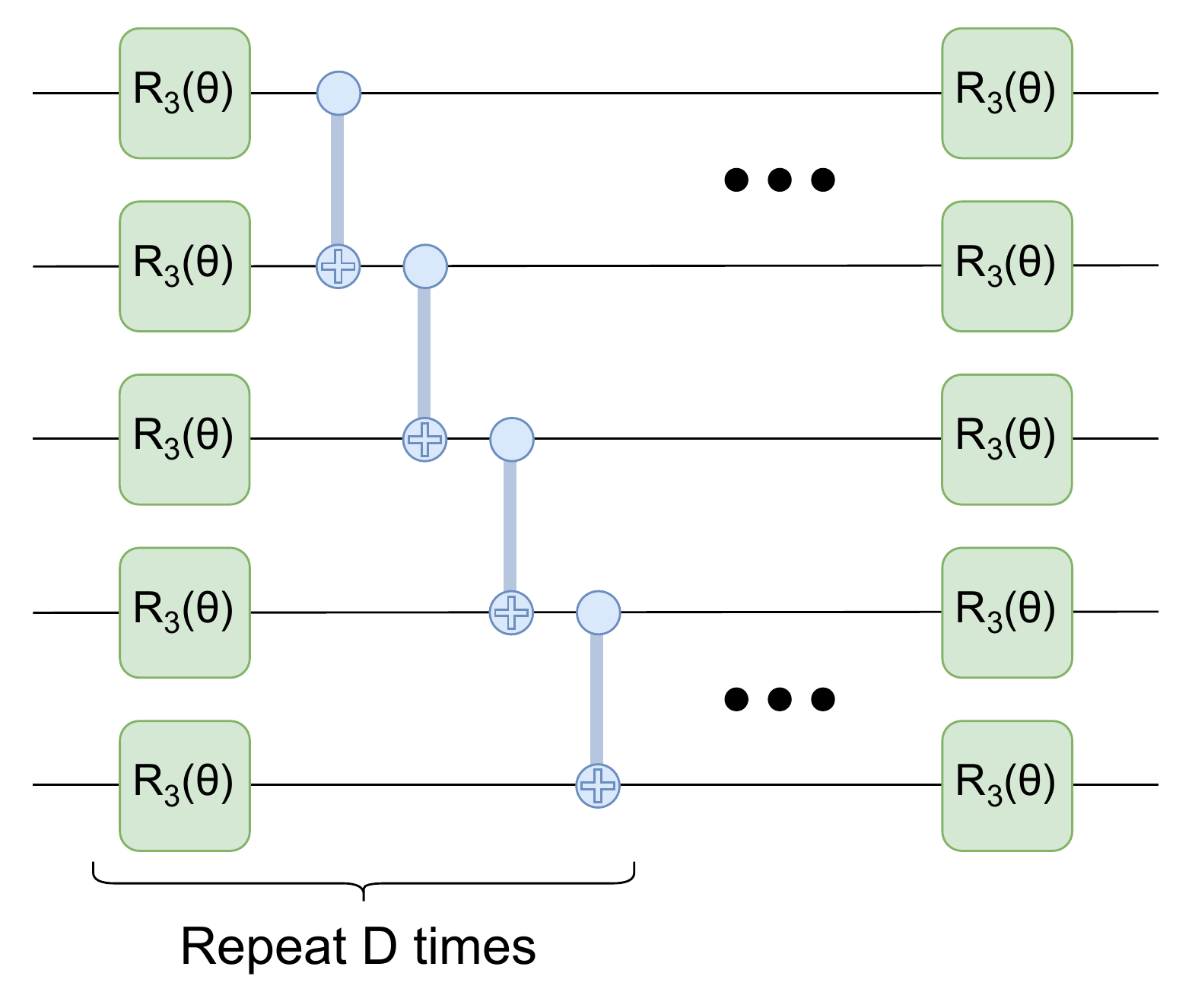} 
\caption{\justifying A circuit for generating a chosen N-qubit states that need to be reconstructed using State Tomography. This generates a pure state in the extended space, which leads to a mixed state for the system when the environment qubits are traced out.}
\label{fig_circuit}
\end{figure} 


\section{Results and Discussion}

In this section, we present the outcomes of quantum state tomography (QST) applied to a random \( n \)-qubit quantum state using both Restricted Boltzmann Machines (RBM) and the Maximum Entropy (MaxEnt) formalism. Previous work has demonstrated that MaxEnt enables the reconstruction of quantum state density matrices by utilizing measured expectation values of various operators as constraints. Whether the set of Hermitian operators is informationally complete or incomplete, MaxEnt consistently yields the least biased mixed quantum state that aligns with the imposed constraints.\cite{gupta2021,gupta2021convergence,gupta2022}

Our goal is to reconstruct a general \( n \)-qubit mixed density matrix, given access to \( k \) observable values. To achieve this, we ensure that the state represented using RBM after training estimates these observables as closely as possible to their actual values while simultaneously maximizing the entropy of the state.  

The proposed technique is highly general and can reconstruct any arbitrarily mixed density matrix using any set of known observables provided by the user. For verification of the protocol in this study, we generate simulated data using the following procedure. We implement a random circuit in an extended space (system + environment) and estimate the expectation values of observables by measuring on an appropriate basis. The random circuit consists of multiple layers of randomly chosen single-qubit unitary gates, interleaved with cascades of CNOT gates, which are repeated several times. We consider a general scenario in which the measured observables can be any arbitrary set. In our experiments, we randomly select Pauli observables chosen from \( P_n \), where \( P \in \{I, X, Y, Z\} \) represents the set of Pauli matrices. We conduct various tests with different sets of observables to validate the algorithm's efficacy.


In this quantum state tomography protocol, we employ classical optimization techniques to fine-tune the parameters \( (\vec{a}, \vec{b}, \vec{W}) \) within the analytical framework of the probability distribution. The objective of this optimization process is to obtain a probability distribution that satisfies the given constraints. As depicted in Figure \ref{fig_training}, we begin from an arbitrary initial state and iteratively optimize the cost function until all constraints are satisfied and the entropy of the state is maximized. As shown in Figure \ref{fig_results}, the reconstructed density matrix obtained through RBM successfully maximizes the entropy across various system sizes while maintaining consistency with the given observables. The total deviation of the estimated expectation values from the target expectation values remains below 0.3, demonstrating the effectiveness of our approach.  




To conclude, we have introduced a hybrid framework that combines the principles of the Maximum Entropy formalism with the expressive and computationally efficient ansatz of Restricted Boltzmann Machines for quantum state tomography. Our method extends the applicability of MaxEnt to scenarios involving tomographically incomplete and non-commuting observables—a class of problems that have remained largely unaddressed in previous approaches. By leveraging RBMs, we circumvent the computational bottlenecks traditionally associated with MaxEnt, enabling scalable reconstruction of both pure and mixed quantum states. The successful numerical validation of our protocol highlights its robustness and versatility, setting the stage for practical deployment in near-term quantum experiments and offering a promising direction for quantum state tomography in larger quantum systems.

{\bf Acknowledgment:}
S.K. would like to acknowledge this material is based upon work supported by the U.S. Department of Energy, Office of Science, Office of Basic Energy Sciences Energy Frontier Research Centers program under Award Number DE-SC0025620.

\bibliography{ref}
\balance

\begin{widetext}
\newpage
\section{Appendix}

\subsection{Derivation of Maximum Entropy State from Lagrangian}

For classical systems, we construct a \textbf{Lagrangian} over the probability distribution:

\begin{equation}
L = \sum_i p_i \ln p_i + \lambda_0 \left[\sum_i p_i -1\right] + \lambda_1 \left[\sum_i p_i O_1(i) - \langle O_1 \rangle \right] + \cdots
\end{equation}

Taking the derivative with respect to \( p_i \) and setting it to zero:

\begin{equation}
\frac{\partial L}{\partial p_i} = \ln p_i + 1 + \lambda_0 + \lambda_1 O_1(i) + \cdots = 0
\end{equation}

Solving for \( p_i \):

\begin{equation}
\boxed{p_i = \frac{1}{Z} \prod_k e^{-\lambda_k O_k(i)}}
\end{equation}

\vspace{0.2in}
\paragraph{\textbf{Extension to Quantum Systems}:}

For quantum systems, the Lagrangian is defined in terms of the \textbf{density matrices ($\hat{\rho}$)} and \textbf{operators ($\hat{O}_i$)}:

\begin{equation}
L = \text{Tr}(\hat{\rho} \ln \hat{\rho}) + \lambda_0 \left[\text{Tr}(\hat{\rho}) -1\right] + \lambda_1 \left[\text{Tr}(\hat{\rho} \hat{O}_1) - \langle \hat{O}_1 \rangle \right] + \cdots
\end{equation}

The quantum Maximum Entropy formalism differs from its classical counterpart because quantum observables that serve as constraints need not commute. In the classical case, the constraints ensuring probabilities are real and non-negative are inherently satisfied and do not further restrict the entropy. In contrast, the quantum case requires the density matrix to be Hermitian and positive semi-definite. The quantum density matrix looks like its classical counterpart, $Z=exp(\lambda_0)$, $\hat{O}_0=\hat{1}$ and the $\lambda_k$'s are the Lagrange multipliers that enforce the constraints, $Tr(\hat{\rho}_{ME} \hat{O}_k) = \langle\hat{O}_k\rangle$,
\[\hat{\rho}_{ME} = Z^{-1} exp \left(-\sum_{k=1} \lambda_k \hat{O}_k \right) \]

The Hermiticity condition implies that the density matrix and the quantum surprisal \(\hat{I} = -\sum_{k=0} \lambda_k \hat{O}_k\) commute. Since both \(\hat{I}\) and \(\hat{\rho}_{ME}\) are Hermitian, they can be diagonalized simultaneously, simplifying the determination of the Lagrange multipliers.  To ensure the density matrix remains Hermitian under variation, we impose  \(\delta \hat{\rho} = i [\delta W, \hat{\rho}]\) for an arbitrary Hermitian matrix \(W\). This introduces an additional variation term in the Lagrangian:  \(\sum_k i \lambda_k \text{Tr}( \delta W [\hat{\rho}, \hat{O}_k] ).\) By the cyclic property of the trace, this term vanishes when \(\hat{\rho}\) commutes with \(\sum_k \lambda_k \hat{O}_k\). Thus, the validity of the density matrix form is contingent on the set of measured observables $\{O_k\}$ to be commutative amongst each other.

\newpage
\subsection{Calculating Renyi entropy using SWAP operator method}

\begin{lemma}
\[
S_2(\rho_A) = - \ln (\text{Tr}(\rho_A^2)) = - \ln (\langle \text{SWAP}_A \rangle).
\]
\end{lemma}

\textit{Proof}:

The action of the SWAP operator over the product of two states across a partition A. 

\begin{align*}
\text{SWAP}_A &\left( \sum_{\alpha_1, \beta_1} C_{\alpha_1, \beta_1} |\alpha_1\rangle|\beta_1\rangle \right) \otimes \left( \sum_{\alpha_2, \beta_2} D_{\alpha_2, \beta_2} |\alpha_2\rangle|\beta_2\rangle \right) \\
&= \sum_{\alpha_1, \beta_1} C_{\alpha_1, \beta_1} \sum_{\alpha_2, \beta_2} D_{\alpha_2, \beta_2} (|\alpha_2\rangle|\beta_1\rangle) \otimes (|\alpha_1\rangle|\beta_2\rangle).
\end{align*}

Therefore, the expectation value of $\text{SWAP}_A$ over copies of the same state state would be 
\begin{align*}
\langle \psi \otimes \psi &| \text{SWAP}_A | \psi \otimes \psi \rangle \\
&= \sum_{\alpha_1, \alpha_2, \beta_1, \beta_2} C_{\alpha_1, \beta_1} \bar{C}_{\alpha_2, \beta_1} C_{\alpha_2, \beta_2} \bar{C}_{\alpha_1, \beta_2} \\
&= \sum_{\alpha_1, \alpha_2} (\rho_A)_{\alpha_1, \alpha_2} (\rho_A)_{\alpha_2, \alpha_1} = \text{Tr}(\rho_A^2),
\end{align*}

where $(\rho_A)_{\alpha_1, \alpha_2} = \sum_{\beta_1} C_{\alpha_1, \beta_1} \bar{C}_{\alpha_2, \beta_1}$ denotes a matrix element of $\rho_A$. 

\begin{figure}[h!]
    \centering
    \includegraphics[width=0.5\linewidth]{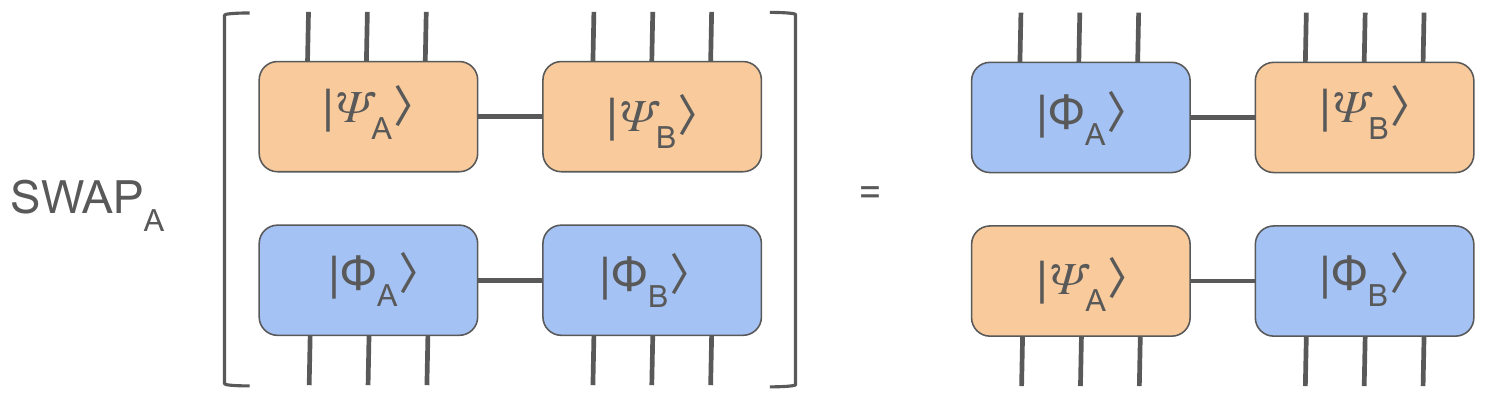}
    \caption{\justifying A tensor network representation of the action of SWAP operator over a product of two states ($|\psi\rangle \otimes |\phi\rangle$) across the region A.}
    \label{fig:SWAP}
\end{figure}

\vspace{0.3in}
\paragraph{\textbf{Calculating $\langle \text{SWAP}_A \rangle$ using sampling }}

\begin{lemma}
\[\langle \text{SWAP}_A \rangle \approx \sum_{u,v \sim |\psi|^2} \left( \frac{\psi(u') \psi(u')}{\psi(u) \psi(v)} \right)\]
where bitstring $\{u,v\}$ are sampled from the distribution $|\psi|^2$, and $\text{SWAP}_A |u,v\rangle = |u',v'\rangle$
\end{lemma}

\textit{Proof:}
\begin{align*}
\langle \psi \otimes \psi &| \text{SWAP}_A | \psi \otimes \psi \rangle \\
&=\sum_{u,v,u''v''} \langle \psi \otimes \psi | u,v \rangle \langle u,v| \text{SWAP}_A | u'', v'' \rangle \langle u'',v''|\psi \otimes \psi\rangle \\
& \qquad \text{Let } \;  SWAP_A |u,v\rangle = |u',v'\rangle\\
&=\sum_{u,v,u''v''} \langle \psi \otimes \psi | u,v \rangle \langle u',v'| u'', v'' \rangle \langle u'',v''|\psi \otimes \psi\rangle \\
&=\sum_{u,v} \langle \psi \otimes \psi | u,v\rangle   \langle u',v'|\psi \otimes \psi\rangle \\
&=\sum_{u,v} \psi^*(u) \psi^*(v) \psi(u') \psi(v') \\
&=\sum_{u,v} |\psi(u)|^2 |\psi(v)|^2 \left( \frac{\psi(u') \psi(u')}{\psi(u) \psi(v)} \right) \\
\end{align*}

Calculating $(u',v')$ is trivial for a set of bitstrings $(u,v)$. One just exchanges the bits in region A amongst the bitstrings. 
Then, computing the coefficients of the wavefunction corresponding to these new bitstrings can be done trivially as the analytical expression of the wavefunction is known. 

\begin{align*}
\langle \text{SWAP}_A \rangle &= \frac{\sum_{u,v} |\psi(u)|^2 |\psi(v)|^2 \left( \frac{\psi(u') \psi(u')}{\psi(u) \psi(v)} \right)}{\sum_{u,v} |\psi(u)|^2 |\psi(v)|^2} \\
&\approx \sum_{u,v \sim |\psi|^2} \left( \frac{\psi(u') \psi(u')}{\psi(u) \psi(v)} \right)
\end{align*}

\newpage
\subsection{Gradient of Cost Function w.r.t RBM parameters}
\textbf{Cost Function}:
\[C = - S(\rho) + \sum_i \xi_i \Big[\langle \hat{O}_i \rangle_\rho - O_i^{target} \Big]^2\]

   where \( \langle O_i \rangle_{\rho} = \text{Tr}(\rho O_i) \) 

\vspace{0.3in}
\textbf{Derivative of the observable term ($C_{obs}$)}:

Apply the chain rule:
\[
\frac{\partial C_{obs}}{\partial x_i} = \sum_i 2\xi_i \left( \langle O_i \rangle_\rho - O_i^{\text{target}} \right) \frac{\partial \langle O_i \rangle}{\partial x_i}
\]

\begin{lemma}
    $$\partial_{x_i} \langle O \rangle = \left\langle \mathcal{D}_{x_i} \odot O^T  \right\rangle_{\rho} - \left\langle \mathcal{D}_{x_i} \right\rangle_{\rho} \left\langle O \right\rangle_{\rho}$$ 
    where the `$\odot$' represents element-wise Hadamard product between matrices.
\end{lemma}

* Refer to Supplementary info of \cite{sajjan2024polynomially} for proof

\begin{lemma}
    $\partial_{x_i} \rho (v,v') = \mathcal{D}_{x_i} (v,v') \odot \rho (v,v')$. The matrix $\mathcal{D}_{x_i} (v,v')$ for various parameters $x_i$ is given as follows:
\end{lemma}

\[
\begin{array}{|c|c|}
\hline
x_i & \mathcal{D}^v_{v_i'} (x_i) \\ 
\hline
\text{Re}(a_k) &  -\beta \left( v_k + v_k' \right) \\ 
\hline
\text{Im}(a_k) &  -i\beta \left( v_k - v_k' \right) \\ 
\hline
\text{Re}(b_p) &  \beta \Big\{ \tanh\left( \beta b_p + \beta \sum_{i=1}^n W_{ip} v_i \right) \\ 
& + \tanh\left( \beta b_p^* + \beta \sum_{i=1}^n W_{ip}^* v_i' \right) \Big\} \\ 
\hline
\text{Im}(b_p) &  i\beta \Big\{ \tanh\left( \beta b_p + \beta \sum_{i=1}^n W_{ip} v_i \right) \\ 
& - \tanh\left( \beta b_p^* + \beta \sum_{i=1}^n W_{ip}^* v_i' \right) \Big\} \\ 
\hline
\text{Re}(W_{kp}) & \beta \Big\{ \tanh\left( \beta b_p + \beta \sum_{i=1}^n W_{ij} v_i \right) v_k  \\ 
& + \tanh\left( \beta b_p^* + \beta \sum_{i=1}^n W_{ip}^* v_i' \right) v_k'  \Big\} \\ 
\hline
\text{Im}(W_{kp}) & i \beta \Big\{ \tanh\left( \beta b_p + \beta \sum_{i=1}^n W_{ip} v_i \right) v_k  \\ 
& - \tanh\left( \beta b_p^* + \beta \sum_{i=1}^n W_{ip}^* v_i' \right) v_k'  \Big\} \\ 
\hline

\end{array}
\]

* Refer to Supplementary info of \cite{sajjan2024polynomially} for proof

\vspace{0.3in}
\textbf{Derivative of the second order Renyi entropy}:
\[S_2 = -ln(\langle SWAP_A \rangle)\]
\[\partial_{x_i} S_2 = \frac{1}{\langle \text{SWAP}_A \rangle} \partial_{x_i} \langle \text{SWAP}_A \rangle\]

\[\langle \text{SWAP}_A \rangle = \frac{\langle \psi \otimes \psi | \text{SWAP}_A | \psi \otimes \psi \rangle}{\langle \psi \otimes \psi| \psi \otimes \psi \rangle} \]

The derivative of the numerator 
\begin{align*}
    \partial_{x_i} &\langle \psi \otimes \psi | \text{SWAP}_A | \psi \otimes \psi \rangle \\
    &= \partial_{x_i} \sum_{u,v} \psi^*(u) \psi^*(v) \psi(u') \psi(v') \\
    &= \partial_{x_i} \sum_{u,v} \rho(u,u') \rho(v,v')  \\
    &= \sum_{u,v}  \Big[\mathcal{D}(u,u') + \mathcal{D}(v,v') \Big] \odot \rho(u,u') \rho(v,v') \\
\end{align*}

The derivative of the denominator 
\begin{align*}
    \partial_{x_i} &\langle \psi \otimes \psi | \psi \otimes \psi \rangle \\
    &= \partial_{x_i} \sum_{u,v} \psi^*(u) \psi^*(v) \psi(u) \psi(v)   \\
    &= \partial_{x_i} \sum_{u,v} \rho(u,u) \rho(v,v)  \\
    &= \sum_{u,v} \Big[\mathcal{D}(u,u) + \mathcal{D}(v,v) \Big] \odot \rho(u,u) \rho(v,v) \\
\end{align*}

\begin{align*}
    \partial_{x_i} \langle \text{SWAP}_A \rangle &= \frac{  \partial_{x_i} \langle \psi \otimes \psi | \text{SWAP}_A | \psi \otimes \psi \rangle }{\langle \psi \otimes \psi| \psi \otimes \psi \rangle} - \frac{ \langle \text{SWAP}_A \rangle \partial_{x_i} \langle \psi \otimes \psi| \psi \otimes \psi \rangle }{\langle \psi \otimes \psi| \psi \otimes \psi \rangle} \\ 
    &= \frac{1}{[Tr(\rho)]^2}\sum_{u,v} \Big[\mathcal{D}(u,u') + \mathcal{D}(v,v') \Big] \odot \rho(u,u') \rho(v,v') \\  
    & \qquad \qquad - \frac{\langle \text{SWAP}_A \rangle}{[Tr(\rho)]^2}  \sum_{u,v}   \Big[\mathcal{D}(u,u) + \mathcal{D}(v,v) \Big] \odot \rho(u,u) \rho(v,v)
\end{align*}

\end{widetext}

\end{document}